# Input-Output Clustering Criterion (IOCC) for Optimizing Distributed Antenna Locations

Zekeriya Uykan, *Senior Member, IEEE,* and Riku Jäntti, *Senior Member*

*Abstract—* In this paper, we propose an input-output space clustering criterion (IOCC) to optimize the locations of the remote antenna units (RAUs) of generalized Distributed Antenna Systems (DASs) under sum power constraint. In IOCC, the input space refers to RAU location space and output space refers to location specific ergodic capacity space for noise-limited environments. Given a location-specific *arbitrary* desired ergodic capacity function over a geographical area, we define the error as the difference between actual and desired ergodic capacity. Our investigations show that *i)* the IOCC provides an upper bound to the cell averaged ergodic capacity error; and *ii)* the derived upper bound is equal to a weighted quantization error function in *location-capacity* space (input-output space) and *iii)* the upper bound can be made arbitrarily small by a clustering process increasing the number of RAUs for a *feasible* DAS. IOCC converts the RAU location problem into a codebook design problem in vector quantization in *input-output* space, and thus includes the Squared Distance Criterion (SDC) for DAS in [15] (and other related papers) as a special case, which takes only the *input* space into account. Computer simulations confirm the theoretical findings and show that the IOCC outperforms the SDC for DAS in terms of the defined cell averaged *"effective"* ergodic capacity.

## I. INTRODUCTION AND MOTIVATION

Distributed Antenna System (DAS) has recently been a hot research area after various works showed that DAS outperforms traditional Co-located Antenna Systems (CASs) in terms of not only transmit power saving but also spectral efficiency for various *outdoor* environments (e.g. [1], [2], [3]). This motivates DAS to be considered as a new cellular communication structure for future wireless communication networks [2]: Unlike the traditional CAS where all antenna elements are co-located, the DAS distributes its antenna elements located at Remote Antenna Units (RAUs) geographically over the cell area. For further information and references about DAS, see e.g. [2]. Although there is a vast literature on optimal location of transmitters for micro/macro-cellular radio communications systems (see e.g. [4],[5],[6],[7],[8],[9],[10],[11]), the works on DAS antenna location optimization have started proliferating only in recent years.

The system performance improvements of DAS in terms of power saving and spectral efficiency highly depend on the locations of its RAUs [15],[2]. Location optimization of the RAUs of DAS, in general, is a very challenging problem due to the high interactions and complexity between the RAU locations and the DAS' spectral capacity. Therefore, instead of optimizing directly a capacity related cost function like the cell averaged ergodic capacity, the authors like in [15],[17],[24] first derive a lower bound to the actual cost function, and then maximize the lower bound, which yields a codebook design problem and which can be solved easily by a clustering process. Some analytically optimal RAU location results are available only for some simplified special cases due to the complexity of the optimization problem. Several papers analyzed the performance of DAS with fixed RAU locations for various transmit strategies for uplink or downlink. The optimal RAU location in terms of "area averaged bit error probability" for linear downlink DAS is derived in [18]. An optimal radius for RAU locations of DAS in circular-layout is investigated in [19]. The authors of [3] propose an iterative algorithm to determine optimal RAU locations based on stochastic approximation theory. So-called "Squared Distance Criterion (SDC)" was proposed in [15] for antenna location determination in Generalized DAS (GDAS) [25] in order to maximize a *lower bound* of the cell averaged ergodic capacity. The paper [15] converts the RAU location problem into the codebook design problem in vector quantization [16]. This implies that any clustering algorithm like Lloyd or *k*-means can be used to optimize the RAU locations of the DAS [15]. As a result, the SDC [15] received much intention within the DAS academic community, and following the footsteps of the SDC and the analysis in [15], several other papers further investigate the SDC for different DAS scenarios: In [17], the SDC is applied to the downlink DAS with Selection Transmission (ST) in a single-cell. Squared-distance-divided-power-criterion is proposed in [24] for linear DAS, which similarly maximizes a *lower bound* to the ergodic capacity. A RAU location design method for single-cell and two-cell downlink DAS with ST is presented in [14] which maximizes a *lower bound* of the expected Signal-to-Noise-Ratio (SNR). The results in [14] are either the same or quite close to the SDC solution. Similarly, an SNR criterion is used for DAS with multiple-antenna ports in [13]. In [12], author extends the SDC results to single and two-cell DAS with ST, Maximal Ratio Transmission (MRT) and zero-forcing beamforming under sum power constraint by maximizing a *lower bound* of the expected SNR.

In all aforementioned works in DAS literature, a performance index like cell averaged ergodic capacity, or expected SNR is optimized evenly over the whole geographical area of the DAS without any location-dependent desired performance preferences. However, in many practical cases, for example, the *desired* ergodic capacity (or desired SNR) depends on location. Some locations/spots/areas may demand more capacities than some other locations/spots/areas for various reasons. Therefore, there is a need for optimizing the RAU locations for cases where location-dependent desired ergodic capacity function is specified. This paper addresses exactly this question, as a first work in DAS literature to our best knowledge. In this paper, we follow a different approach than any others mentioned above, and propose an input-output space clustering criterion (IOCC) to optimally determine the locations of the RAUs of GDAS taking also location-dependent desired ergodic capacity function into account: For a given *arbitrary* mobile user location distribution and *arbitrary* location-specific ergodic capacity function in the geographical area of GDAS, what are the optimum RAU locations? Our investigations, partly inspired by the analysis in [20],[21],[22] show that *i)* the IOCC converts the RAU location problem into codebook design problem in the location-capacity space (input-output space), and therefore includes the SDC for DAS in [15] as its special case, which considers only location space (input space). *ii)* the IOCC provides an *upper bound* to the ergodic capacity error where capacity error is defined as the difference between the (location-dependent) desired and actual ergodic capacity. *iii)* The derived *upper bound* can be made arbitrarily small by increasing the number of RAUs.

The paper is arranged as follows: We present the system model in section II. The proposed IOCC is presented and analyzed in section III. Simulation results are shown in section IV, followed by the conclusions in section V.

*Notation:* Throughout the paper, bold upper and bold lower case letters denote matrices and vectors, respectively, and superscript $(.)^T$ denotes transpose. $\mathbf{I}_M$ is the $M \times M$ identity matrix, $\mathbf{1}_M$ is a column vector of 1's, the operation $\otimes$ shows entry-wise multiplication, and $E\{\cdot\}$ represents the expectation.

## II. SYSTEM MODEL

Let's consider a GDAS [25] with *K* RAUs in each cell and *M* antenna elements in each RAU, and every user terminal have one antenna. We examine noise-limited environment as in [15]. This corresponds to an isolated cell case or any frequency reuse case where the co-channel interference is small compared to the thermal noise. If the RAU includes multiple co-located antenna elements, then the channels between one RAU and the user undergo the same large-scale fading. All channels between the antennas and the user are assumed to be flat fading and slow fading. Let's denote the channel vector from the *n*'th RAU to the user as

$$\mathbf{h}_k = \sqrt{\frac{s_k}{d_k^\alpha}} [h_{k,1} \; h_{k,2} \cdots h_{k,M}]^T \qquad (1)$$

where $\alpha$ is the path loss exponent, $s_k$ is the large-scale fading (e.g. shadow fading) term (between the user and the *k*'th RAU) and is modeled as log-normal random variable (i.e., $10\log_{10}(s_k)$ is a zero mean Gaussian random variable (rv) with standard deviation $\sigma_s$), and $h_{k,m}$ (*k=1,...,K, m=1,...,M*) represents small-scale fading (multipath) (e.g. Rayleigh fading) term (between the user and the *m*'th antenna element of the *n*'th RAU), and is modeled as a unit-variance circularly symmetric complex Gaussian rv. Large-scale and small-scale fadings are independent. Then the $NM \times 1$ dimensional channel vector **h** between the user and the DAS has the form

$$\mathbf{h} = [\mathbf{h}_1^T \ \mathbf{h}_2^T \ \cdots \ \mathbf{h}_K^T]^T \qquad (2)$$

Representing the transmit power of the $n$'th RAU as $\mathbf{p}_k = p_k \mathbf{1}_M^T$, $k=1,2,\ldots,K$, we denote the complete transmit power vector as

$$\mathbf{p} = [\mathbf{p}_1^T \ \mathbf{p}_2^T \ \cdots \ \mathbf{p}_K^T]^T \qquad (3)$$

Then the transmitted signal vector $\mathbf{y}$ from DAS to the user can be written as

$$\mathbf{y} = (\mathbf{p} \otimes \mathbf{h})a + \boldsymbol{\varsigma} \qquad (4)$$

where $a$ is the transmitted symbol, and channel vector $\mathbf{h}$ and transmit power vector $\mathbf{p}$ is defined in (2) and (3), respectively, and $\boldsymbol{\varsigma}$ is a zero-mean complex additive white Gaussian noise vector whose covariance matrix is $\mathrm{E}\{\boldsymbol{\varsigma}\boldsymbol{\varsigma}^H\} = \sigma_\varsigma \mathbf{I}_{NL}$ where $\sigma_\varsigma > 0$. Instantaneous downlink SNR in case of the MRT [23] or Maximal Ratio Combining (MRC) for an arbitrary user location $\mathbf{x}_l$ is equal to

$$\theta_{h,s,\mathbf{x}_l} = \frac{1}{\sigma_\varsigma^2} \sum_{k=1}^{N} \frac{p_k s_k g_k}{\|\mathbf{x}_l - \mathbf{c}_k\|_2^\alpha} \qquad (5)$$

where $k$ is the index of the RAU and thus $g_k = \sum_{m=1}^{M} |h_{k,m}|^2$, and $p_k$ is the transmit power of the RAU $n$, where $n \in \{1,\cdots,N\}$.

In the design of practical wireless systems, different parts of the cell area may demand different desired capacities (e.g. [6]). This is due to the user location distribution over the whole cell area, and due to some geographical constraints, or network cost constraints etc. For example, the desired ergodic capacity in hot spot areas like school campuses, meeting areas, etc is much higher than those in remote and less densely populated areas. So, the desired capacity depends on the location. In practice, naturally there is a minimum distance that should be kept between the user location and any RAU location. Let's denote this minimum distance as $d_{min}$. Thus, the system performance of the GDAS should be calculated for

$$\overline{\Omega}: \ \mathbf{x} \in \Omega \ \ni \ \|\mathbf{x} - \mathbf{c}_k\|_2 \geq d_{min}, \ k=1,\cdots,N \qquad (6)$$

where $\Omega$ denotes the whole geographical area of the GDAS, and $\bar{\Omega}$ represents the sum of all geographical areas satisfying the minimum RAU-user distance in (6). Let the location vector be $\mathbf{x} \in \Re^{q \times 1}$, and the probability distribution function (pdf) of the user location be denoted by $f(\mathbf{x})$. Similarly, the locations of the RAUs $\{\mathbf{c}_k\}_{k=1}^{K} \in \Re^{q \times 1}$. And let $q \times K$ dimensional matrix $\mathbf{C} \in \Re^{q \times K}$ represent a matrix whose columns are the locations of the RAUs $\{\mathbf{c}_k\}_{k=1}^{K}$. The demanded/desired and supplied/actual ergodic capacities conditioned on user position $\mathbf{x}$ is defined as $\Gamma_d(\mathbf{x}) = E_{h,s}\{\log_2(1+\theta_d(\mathbf{x}))\}$, and $\Gamma_a(\mathbf{x}) = E_{h,s}\{\log_2(1+\theta_a(\mathbf{p},\mathbf{C},\mathbf{x}))\}$, where $\theta_d(\mathbf{x})$ and $\theta_a(\mathbf{p},\mathbf{C},\mathbf{x})$ is desired and actual instantaneous SNRs at location $\mathbf{x}$, respectively, given RAU location matrix $\mathbf{C}$, and RAU transmit power vector $\mathbf{p}$, and $E_{h,s}\{\}$ denotes the expectation with respect to $h$ (small-scale fading) and $s$ (large-scale fading).

In this paper, we assume that every location/spot $\mathbf{x}$ demands a certain amount of maximum capacity, and any higher supplied capacity exceeding the demanded capacity level will be useless.

*Definition: Wasted ergodic capacity for location* $\mathbf{x}$: If the supplied/actual capacity is more than the demanded/desired capacity for location $\mathbf{x}$, i.e., $\Gamma_d(\mathbf{x}) < \Gamma_a(\mathbf{x})$, then there will be some useless excessive capacity, which we define as "wasted capacity" for that location. So, the wasted capacity for location $\mathbf{x}$ is equal to $\Gamma_{wasted}(\mathbf{x}) = \max\{(\Gamma_a(\mathbf{x}) - \Gamma_d(\mathbf{x})), 0\}$. The defined "wasted capacity" concept is sketched in the schema in Fig. 8.

*Definition: Effective ergodic capacity*: The "effective ergodic capacity" for location $\mathbf{x}$, denoted as $\Gamma_{eff}(\mathbf{x})$, is defined as the amount of capacity which is completely utilized (consumed) by the users (and not wasted) according to a given arbitrary demanded/desired ergodic capacity $\Gamma_d(\mathbf{x})$: So, $\Gamma_{eff}(\mathbf{x}) = \min(\Gamma_d(\mathbf{x}), \Gamma_a(\mathbf{x}))$. Cell averaged *effective* ergodic capacity is then equal to $E_{\mathbf{x}}[\Gamma_{eff}(\mathbf{x})]$.

The RAU location optimization problem may be defined in terms of maximizing the cell-averaged ergodic capacity, denoted as $J_O(\mathbf{p},\mathbf{C})$, under sum power constraints as in e.g. [12],[14],[15]:

$$\max_{\mathbf{p},\mathbf{C}} (J_O(\mathbf{p},\mathbf{C}) = E_\mathbf{x}[\Gamma_a(\mathbf{p},\mathbf{C},\mathbf{x})]) \qquad (7)$$
$$\text{s.t. } \|\mathbf{p}\|_1 \leq p_{tx}^{sum}$$

where $p_{tx}^{sum}$ is the maximum total DAS transmit power. Denoting the cell-averaged demanded ergodic capacity as $\hat{\Gamma}_d = E_\mathbf{x}[\Gamma_d(\mathbf{x})] + \delta$, where $\delta \geq 0$, the maximization problem may be turned into a minimization problem under the sum power constraint as follows:

$$\max_{\mathbf{p},\mathbf{C}} (J_O(\mathbf{p},\mathbf{C})) = \min_{\mathbf{p},\mathbf{C}} (-J_O(\mathbf{p},\mathbf{C})) \qquad (8)$$
$$= \min_{\mathbf{p},\mathbf{C}} (\hat{\Gamma}_d - J_O(\mathbf{p},\mathbf{C}))$$
$$\text{s.t. } \|\mathbf{p}\|_1 \leq p_{tx}^{sum}$$

Writing $\hat{\Gamma}_d = E_\mathbf{x}[\Gamma_d(\mathbf{x})] + \delta$ and $J_O(\mathbf{p},\mathbf{C})$ in (8), the function to be minimized is obtained as

$$J_U(\mathbf{p},\mathbf{C}) = \hat{\Gamma}_d - J_O(\mathbf{p},\mathbf{C}) = E_\mathbf{x}[\Gamma_d(\mathbf{x}) + \delta - \Gamma_a(\mathbf{p},\mathbf{C},\mathbf{x})] \qquad (9)$$

where $\delta \geq 0$. From (8)-(9), the desired/demanded ergodic capacity function $\Gamma_d(\mathbf{x})$ has no effect on the solution of the minimization $J_U(\mathbf{p},\mathbf{C})$ in (9). However, in this paper we aim to maximize the cell-averaged ergodic capacity according to the provided arbitrary $\Gamma_d(\mathbf{x})$, which would imply for example that the supplied ergodic capacity is supposed to be higher in those areas/locations/spots where the demanded ergodic capacity is higher. Therefore, we need to take the absolute value of the argument of the expectation in (9), which gives an upper bound to the $J_U(\mathbf{p},\mathbf{C})$ in (9), and we obtain a generic cost function, denoted as $J_{E,1}(\mathbf{p},\mathbf{C})$, under the sum power constraint as follows:

$$J_U(\mathbf{p},\mathbf{C}) \leq J_{E,1}(\mathbf{p},\mathbf{C}) = E_\mathbf{x}[|\Gamma_d(\mathbf{x}) + \delta - \Gamma_a(\mathbf{p},\mathbf{C},\mathbf{x})|] \qquad (10)$$
$$\text{s.t. } \|\mathbf{p}\|_1 \leq p_{tx}^{sum}$$

The function $J_{E,1}(\mathbf{p},\mathbf{C})$ in (10) to be minimized is a generic cost function because

1. minimizing $J_{E,1}(\mathbf{p},\mathbf{C})$ with $\delta=0$ in (10) is equal to maximizing the defined *effective* ergodic capacity while at the same time minimizing the wasted ergodic capacity, i.e.

$$\min J_{E,1}(\mathbf{p},\mathbf{C}) = \left(\max E_\mathbf{x}\left[\Gamma_{eff}(\mathbf{x})\right] \text{ and } \min E_\mathbf{x}\left[\Gamma_{wasted}(\mathbf{x})\right]\right) \quad (11)$$
$$\text{s.t. } \|\mathbf{p}\|_1 \leq p_{tx}^{sum}$$

because $\Gamma_{eff}(\mathbf{x}) = \min(\Gamma_d(\mathbf{x}),\Gamma_a(\mathbf{x}))$ and

$\Gamma_{wasted}(\mathbf{x}) = \max\{(\Gamma_a(\mathbf{x})-\Gamma_d(\mathbf{x})), 0\}$.

2. minimizing $J_{E,1}(\mathbf{p},\mathbf{C})$ with $\Gamma_d(\mathbf{x})+\delta \geq \Gamma_a(\mathbf{p},\mathbf{C},\mathbf{x})$, $\forall \mathbf{x}$, is equal to maximizing the original cell-averaged ergodic capacity, i.e.,

$$\min J_{E,1}(\mathbf{p},\mathbf{C}) = \max J_O(\mathbf{p},\mathbf{C}) \quad (12)$$
$$\text{s.t. } \|\mathbf{p}\|_1 \leq p_{tx}^{sum}$$

which is an immediate result from (8) and (10).

It's worth to emphasize that the location-specific desired/demanded ergodic capacity function $\bar{\theta}_d(\mathbf{x})$ is arbitrary in the formulation.

Defining ergodic capacity error at location $\mathbf{x}$ as

$$e(\mathbf{x}) = \Gamma_d(\mathbf{x}) + \delta - \Gamma_a(\mathbf{p},\mathbf{C},\mathbf{x}) \quad (13)$$

where $\delta \geq 0$, eq.(10) can be re-written as

$$\min_{\mathbf{p},\mathbf{C}}\{J_{E,1}(\mathbf{p},\mathbf{C}) = E_\mathbf{x}[e(\mathbf{x})]\} \quad (14)$$
$$\text{s.t. } \|\mathbf{p}\|_1 \leq p_{tx}^{sum}$$

Although $J_{E,1}(\mathbf{p},\mathbf{C})$ in (10) and (14) is a generic and meaningful in practice as explained in (11)-(12), its main "deficiency" is the fact that the $\Gamma_d(\mathbf{x})$ has no effect on the minimization of $J_{E,1}(\mathbf{p},\mathbf{C})$ when

$\Gamma_d(\mathbf{x}) + \delta \geq \Gamma_a(\mathbf{p}, \mathbf{C}, \mathbf{x})$, $\forall \mathbf{x}$, as seen from (8)-(9), and (10). This is against our aim in this paper, because, as mentioned earlier, we aim to design a DAS such that its location-specific supplied/actual ergodic capacity function $\Gamma_a(\mathbf{p}, \mathbf{C}, \mathbf{x})$ resembles the given demanded ergodic capacity function $\Gamma_d(\mathbf{x})$ while still maximizing the cell-averaged ergodic capacity $E_\mathbf{x}[\Gamma_a(\mathbf{p}, \mathbf{C}, \mathbf{x})]$. Thus the absolute capacity error in (14) can not provide the intended solution for the case $\Gamma_d(\mathbf{x}) + \delta \geq \Gamma_a(\mathbf{p}, \mathbf{C}, \mathbf{x})$, i.e., $e(\mathbf{x}) \geq 0$, $\forall \mathbf{x}$. However, taking squared error instead of the absolute error in (14) solves the issue when $e(\mathbf{x}) \geq 0$, $\forall \mathbf{x}$, because the error squares makes it sure that not only the $\Gamma_a(\mathbf{p}, \mathbf{C}, \mathbf{x})$ is shaped according to the given $\Gamma_d(\mathbf{x})$ but also the $\Gamma_a(\mathbf{p}, \mathbf{C}, \mathbf{x})$ is maximized for $\forall \mathbf{x}$. Therefore, we also need to introduce the following cost function

$$\min_{\mathbf{p},\mathbf{C}} \left\{ J_{E,2}(\mathbf{p},\mathbf{C}) = E_\mathbf{x}\left[e^2(\mathbf{x})\right] \right\} \quad (15)$$
$$\text{s.t. } \|\mathbf{p}\|_1 \leq p_{tx}^{sum}$$

which is minimization of the cell averaged ergodic capacity error squares under the sum power constraint. Comparing (14) and (15) for the cases where $e(\mathbf{x})$ may be positive and negative, the $J_{E,2}(\mathbf{p},\mathbf{C})$ in (15) gives more emphasis to those locations where higher capacities are desired than the $J_{E,1}(\mathbf{p},\mathbf{C})$ in (14) does. Furthermore, it's well known that minimizing the error squares is mathematically more tractable than the absolute errors.

Our investigations in section III show that minimizing $J_{E,1}(\mathbf{p},\mathbf{C})$ in (14) and $J_{E,2}(\mathbf{p},\mathbf{C})$ in (15) gives similar results: The analysis yields upper bounds to the cell averaged ergodic capacity error functions in (14) and (15) and these upper bounds can be made arbitrarily small by increasing the number of RAUs. This implies that we can arbitrarily get closer to the maximum effective ergodic capacity for a *feasible* DAS.

In brief, the RAU location problem in GDAS is defined as follows: For a given *arbitrary* user location distribution $f(\mathbf{x})$ and location-dependent desired ergodic capacity function $\Gamma_d(\mathbf{x})$, what are the optimum RAU locations minimizing *i*) the $J_{E,1}(\mathbf{C})$ in (14) and *ii*) the $J_{E,2}(\mathbf{C})$ in (15) under the sum power constraint?

III. INPUT-OUTPUT CLUSTERING CRITERION (IOCC) FOR DETERMINING GDAS RAU LOCATIONS

*A. IOCC in Statistical Setting*

The SNR in (5) is instantaneous SNR. The average SNR for a particular user location **x**, denoted by $\bar{\theta}_a(\mathbf{x})$, is obtained by averaging it over the small-scale and large-scale fadings:

$$\bar{\theta}_a(\mathbf{x}) = \mathrm{E}_{h,s}\{\theta_{h,s,\mathbf{x}_l}\} = \frac{1}{\sigma_\varsigma^2} \sum_{k=1}^{K} p_k \mathrm{E}_{h,s}\{s_k g_{k,\mathbf{x}}\} \phi(\|\mathbf{x}_l - \mathbf{c}_k\|_2) \quad (16)$$

where $\mathbf{x} \in \bar{\Omega}$, $s_k$ represents the (RAU specific) large-scale fading, $g_{k,\mathbf{x}}$ shows the (location specific) sum of small-scale fading, as explained above, and $\phi(\|\mathbf{x}_l - \mathbf{c}_k\|_2) = 1/\|\mathbf{x}_l - \mathbf{c}_k\|_2^\alpha$, in which $\alpha$ is the path loss exponent (typically $2 \leq \alpha \leq 6$). For any location $\mathbf{x} \in \Omega$, let's denote the RAU location which is the closest to **x** as $\mathbf{c}_{k(\mathbf{x})}$, where index $k(\mathbf{x}) \in \{1,2,\cdots,K\}$. In other words, $\mathbf{c}_{k(\mathbf{x})}$ is the one which satisfies $\|\mathbf{c}_{k(\mathbf{x})} - \mathbf{c}_k\| = \min\{\|\mathbf{x} - \mathbf{c}_k\|\}_{k=1}^{K}$.

*1) Location Optimization using $J_{E,1}(\mathbf{C})$ in (14)*

In this subsection, our aim is to derive an upper bound to $J_{E,1}(\mathbf{p},\mathbf{C})$ in (14): Applying the mean-value theorem to the function $\log_2(\cdot)$ in (14), we obtain an upper bound (UB)

$$J_{E,1}(\mathbf{p},\mathbf{C}) = \mathrm{E}_\mathbf{x}[\Gamma_d(\mathbf{x}) + \delta - \Gamma_a(\mathbf{x})] \quad (17)$$
$$\leq \mathrm{E}_\mathbf{x}\mathrm{E}_{h,s}[\upsilon|\theta_d(\mathbf{x}) - \theta_a(\mathbf{p},\mathbf{C},\mathbf{x})| + \delta]$$

where $\upsilon$ is the global Lipschitz constant of the function $\log_2(\cdot)$ for the area $\bar{\Omega}$. So, $\upsilon = \max\{1/(1+\min(\theta_d(\mathbf{x}),\theta_a(\mathbf{x})))\}$, and thus $0 < \upsilon < 1$. Performing the expectation $\mathrm{E}_{h,s}\{\}$ yields

$$J_{E,1}(\mathbf{p},\mathbf{C}) \leq \mathrm{E}_\mathbf{x}[\upsilon|\bar{\theta}_d(\mathbf{x}) - \bar{\theta}_a(\mathbf{p},\mathbf{C},\mathbf{x})| + \delta] \quad (18)$$

where $\bar{\theta}_d(\mathbf{x})$ is desired average SNR at location **x**, and actual average SNR $\bar{\theta}_a(\mathbf{x})$ is given by (16). Both $\bar{\theta}_d(\mathbf{x})$ and $\bar{\theta}_a(\mathbf{x})$ are averaged over the very same small and large-scale fadings because the RAU locations and user location **x** and the wireless environment are all the same for both. Adding and subtracting the term

$$\bar{\theta}_a(\mathbf{C},\mathbf{c}_{k(\mathbf{x})}) = \frac{1}{\sigma_\varsigma^2}\sum_{n=1}^{K} p_n E_{h,s}\{s_n g_{n,\mathbf{c}_{k(\mathbf{x})}}\}\phi(d_{k,n}), \text{ where } \phi(d_{k,n}) = \min\{d_{\min}^{-\alpha}, \|\mathbf{c}_{k(\mathbf{x})} - \mathbf{c}_n\|_2^{-\alpha}\} \text{ due to the minimum user-RAU}$$

distance requirement, into the argument of the integral in (18) gives

$$J_{E,1}(\mathbf{p},\mathbf{C}) \le \upsilon \int_{\mathbf{x}\in\Omega} \left( \left| \begin{array}{c} \bar{\theta}_d(\mathbf{x}) - \bar{\theta}_a(\mathbf{C},\mathbf{x}) + \delta/\upsilon \\ + \bar{\theta}_a(\mathbf{C},\mathbf{c}_{k(\mathbf{x})}) - \bar{\theta}_a(\mathbf{C},\mathbf{c}_{k(\mathbf{x})}) \end{array} \right| \right) f(\mathbf{x}) d\mathbf{x} \qquad (19)$$

The argument of the integral is always positive due to the absolute function. Thus, (19) is upper bounded by

$$J_{E,1}(\mathbf{p},\mathbf{C}) \le \upsilon \int_{\mathbf{x}\in\Omega} \left( \begin{array}{c} |\bar{\theta}_a(\mathbf{C},\mathbf{x}) - \bar{\theta}_a(\mathbf{C},\mathbf{c}_{k(\mathbf{x})})| \\ + |\bar{\theta}_d(\mathbf{x}) + \delta/\upsilon - \bar{\theta}_a(\mathbf{C},\mathbf{c}_{k(\mathbf{x})})| \end{array} \right) f(\mathbf{x}) d\mathbf{x} \qquad (20)$$

It's assumed that large and small-scale fading random variables $s_n$ and $g_{n,\mathbf{x}}$ are independent, and average large-scale fading $E_s\{s_n\}$ is RAU location-specific and therefore is same for locations $\mathbf{x}$ and $\mathbf{c}_k$. In the Appendix, we prove that the average SNR function $\bar{\theta}_a(\cdot)$ in (16) has a global Lipschitz constant $\nu_{glob}$ for the interval $[d_{\min},\infty)$, i.e.:

$$\|\bar{\theta}_a(\mathbf{x}_i) - \bar{\theta}_a(\mathbf{x}_j)\|_1 \le \nu_{glob}\|\mathbf{x}_i - \mathbf{x}_j\|_2, \quad \forall \mathbf{x}_i,\mathbf{x}_j \in \bar{\Omega} \qquad (21)$$

where $\nu_{glob} = \alpha\gamma\left(\sum_{k=1}^{K} p_k \bar{s}_k\right)/\left(\sigma_\varsigma^2 d_{\min}^{(\alpha+1)}\right)$, in which $\alpha$ is the path loss exponent, $\gamma$ is related to the average small-scale fading as defined in (50) in the Appendix, $p_k$ is the transmit power of the $k$'th RAU, $\bar{s}_k$ is the average large-scale fading coefficient related to the $k$'th RAU, $\sigma_\varsigma^2$ is the average noise power, and $d_{\min}$ is the minimum distance between user location and any RAU. So, writing (21) in (20) gives a new UB

$$J_{E,1}(\mathbf{p},\mathbf{C}) \le \upsilon \int_{\mathbf{x}\in\bar{\Omega}} \left( \begin{array}{c} \nu_{glob}\|\mathbf{x} - \mathbf{c}_{k(\mathbf{x})}\|_2 + \\ |\bar{\theta}_d(\mathbf{x}) - \bar{\theta}_a(\mathbf{C},\mathbf{c}_{k(\mathbf{x})}) + \delta/\upsilon| \end{array} \right) f(\mathbf{x}) d\mathbf{x} \qquad (22)$$

and using the fact that the $l_1$-norm of a vector is equal to or greater than its $l_2$-norm, we have

$$J_{E,1}(\mathbf{p},\mathbf{C}) \le \upsilon \int_{\mathbf{x}\in\bar{\Omega}} \left( \left\| \begin{array}{c} \nu_{glob}(\mathbf{x} - \mathbf{c}_{k(\mathbf{x})}) \\ \bar{\theta}_d(\mathbf{x}) - \bar{\theta}_a(\mathbf{c}_{k(\mathbf{x})}) + \delta/\upsilon \end{array} \right\|_1 \right) f(\mathbf{x}) d\mathbf{x} \qquad (23)$$

From (13), (14) and (23) we obtain

$$J_{E,1}(\mathbf{p},\mathbf{C}) \leq UB_{E,1}(\mathbf{p},\mathbf{C}) \qquad (24)$$

where

$$UB_{E,1}(\mathbf{p},\mathbf{C}) = \upsilon v_{glob} \int_{\mathbf{x}\in\overline{\Omega}} \left\| \mathbf{x} - \mathbf{c}_{k(\mathbf{x})} \right\|_1 f(\mathbf{x}) d\mathbf{x} \qquad (25)$$
$$+ \upsilon \int_{\mathbf{x}\in\overline{\Omega}} \left( \overline{\theta}_d(\mathbf{x}) - \overline{\theta}_a(\mathbf{c}_{k(\mathbf{x})}) + \delta/\upsilon \right) f(\mathbf{x}) d\mathbf{x}$$

where $\upsilon$ is the global Lipschitz constant of the function $\log_2(\cdot)$ in area $\overline{\Omega}$. The minimization of (14) is under sum power constraint, i.e. the sum of all RAU's transmit powers is not greater than a total maximum transmit power $p_{tx}^{sum}$. The total transmit power $p_{tx}^{sum}$ can evenly be distributed over the RAUs such that the every RAU would have the same transmit power. Or alternatively, the transmit powers of RAUs can also be optimized for any given location matrix $\mathbf{C}$ in order to further minimize the cell-averaged ergodic capacity error $J_{E,1}(\mathbf{p},\mathbf{C})$. The smaller the $J_{E,1}(\mathbf{p},\mathbf{C})$ the higher the cell-averaged (*effective*) ergodic capacity itself $E_\mathbf{x}\{\Gamma_a(\mathbf{x})\}$, as explained in Section II. Therefore, in what follows we consider the optimum transmit power vector denoted as $\mathbf{p}_{glob}^{opt}$, which globally minimizes $J_{E,1}(\mathbf{p},\mathbf{C})$ under the sum power constraint and over the whole area $\{\mathbf{x},\theta_d(\mathbf{x})\} \ni \mathbf{x}\in\overline{\Omega}$, for any given RAU location matrix $\mathbf{C}$. This implies that for a given matrix $\mathbf{C}$, $J_{E,1}(\mathbf{p}_{glob}^{opt},\mathbf{C}) \leq J_{E,1}(\mathbf{p},\mathbf{C})$ for any $\mathbf{p}$.

*Proposition 1*: Considering the optimum transmit power vector $\mathbf{p}_{glob}^{opt}$, i.e., $J_{E,1}(\mathbf{p}_{glob}^{opt},\mathbf{C}) \leq J_{E,1}(\mathbf{p},\mathbf{C})$ for any $\mathbf{p}$, the upper bound $UB_{E,1}(\mathbf{p}=\mathbf{p}_{glob}^{opt},\mathbf{C})$ in (25) to the ergodic capacity error can be decreased by a codebook design process in vector quantization in *location-SNR/capacity* space.

*Proof:* Defining augmented weighted vectors $\mathbf{x}^{l/O} = [\omega_1 \mathbf{x}^T \; \omega_2(\overline{\theta}_d(\mathbf{x})+\delta)]^T$ and codebook vectors $\boldsymbol{\mu}_{k(\mathbf{x}^{l/O})}^{l/O} = [\omega_1 \boldsymbol{\mu}_{k(\mathbf{x}^{l/O})}^x \; \omega_2 \mu_{k(\mathbf{x}^{l/O})}^\theta]^T$ in *location-SNR/capacity* space, where $\omega_1, \omega_2 > 0$, we consider a codebook design process in vector quantization which minimizes the following quantization error

$$Q_{E,1}(\boldsymbol{\mu}_1^{I/O}, \cdots, \boldsymbol{\mu}_K^{I/O}) = E\left\{\left\|\mathbf{x}^{I/O} - \boldsymbol{\mu}_{k(\mathbf{x}^{I/O})}^{I/O}\right\|_1\right\} \qquad (26)$$

$$= \int_{\mathbf{x} \in \Omega} \left( \begin{array}{l} \omega_1 \left\|\mathbf{x} - \boldsymbol{\mu}_{k(\mathbf{x}^{I/O})}^{x}\right\|_1 + \\ \omega_2 \left|\bar{\theta}_d(\mathbf{x}) + \delta/\upsilon - \mu_{k(\mathbf{x}^{I/O})}^{\theta}\right| \end{array} \right) f(\mathbf{x}) d\mathbf{x}$$

where $k(\mathbf{x}^{I/O})$ shows the index of the cluster to which $\mathbf{x}^{I/O}$ belongs, $\bar{\theta}_d(\mathbf{x})$ is the location-dependent desired SNR function, $f(\mathbf{x})$ is user location distribution, $\boldsymbol{\mu}_{k(\mathbf{x}^{I/O})}^{I/O}$ represents the codebook vector of the cluster $k$ to which the augmented vector $\mathbf{x}^{I/O}$ belongs to in *location-(SNR/capacity)* space. The desired average SNR $\bar{\theta}_d(\mathbf{x})$ in (26) corresponds to the given desired ergodic capacity function $\Gamma_d(\mathbf{x})$. This is because although $\log_2(\cdot)$ is a concave function, and due to the Jensens's inequality, $\log_2(1+\bar{\theta}(\mathbf{x})) \geq \Gamma_d(\mathbf{x})$, the difference $(\log_2(1+\bar{\theta}(\mathbf{x})) - \Gamma_d(\mathbf{x}))$ is too small for the natural logarithm function and thus can be omitted in practical cases. In other words, for any given $\Gamma_d(\mathbf{x})$, there exists a small number $\varepsilon > 0$ such that $\log_2(1+\bar{\theta}_d(\mathbf{x})) \geq \Gamma_d(\mathbf{x}) \geq \log_2(1+\bar{\theta}_d(\mathbf{x})) - \varepsilon$, and therefore, for a given arbitrary $\Gamma_d(\mathbf{x})$, we may assume that there exists a corresponding $\bar{\theta}_d(\mathbf{x})$.

Let's suppose that optimum codebook vectors are found by an optimum codebook design process minimizing (26). Let's take the RAU locations as $\{\mathbf{c}_k = \boldsymbol{\mu}_k^x\}_{k=1}^K$, and let's consider the optimum RAU transmit power vector $\mathbf{p}$ for the codebook vectors $\{\mathbf{c}_k\}_{k=1}^K$, and corresponding $\{\mu_k^\theta\}_{k=1}^K$ in SNR/capacity space. Writing the SNR values $\{\mu_k^\theta\}_{k=1}^K$ in matrix form, taking also the minimum user-RAU distance requirement into account, gives

$$\mathbf{A}_{K \times K} \begin{bmatrix} \bar{s}_1 p_1 \\ \bar{s}_2 p_2 \\ \vdots \\ \bar{s}_K p_K \end{bmatrix} = \begin{bmatrix} \mu_1^\theta \\ \mu_2^\theta \\ \vdots \\ \mu_K^\theta \end{bmatrix} \qquad (27)$$

where

$$\mathbf{A} = \begin{bmatrix} \bar{g}_{1,\mathbf{c}_1}\phi(d_{min}) & \bar{g}_{2,\mathbf{c}_1}\phi(\|\mathbf{c}_1 - \mathbf{c}_2\|_2) & \cdots & \bar{g}_{K,\mathbf{c}_1}\phi(\|\mathbf{c}_1 - \mathbf{c}_K\|_2) \\ \bar{g}_{1,\mathbf{c}_2}\phi(\|\mathbf{c}_2 - \mathbf{c}_1\|_2) & \bar{g}_{2,\mathbf{c}_2}\phi(d_{min}) & \cdots & \bar{g}_{K,\mathbf{c}_2}\phi(\|\mathbf{c}_2 - \mathbf{c}_K\|_2) \\ \vdots & & \ddots & \vdots \\ \bar{g}_{1,\mathbf{c}_K}\phi(\|\mathbf{c}_K - \mathbf{c}_1\|_2) & \bar{g}_{2,\mathbf{c}_K}\phi(\|\mathbf{c}_K - \mathbf{c}_2\|_2) & & \bar{g}_{K,\mathbf{c}_K}\phi(d_{min}) \end{bmatrix}_{K \times K} \quad (28)$$

in which $\bar{g}_{i,\mathbf{c}_j}$ is the average small-scale fading coefficient at location $\mathbf{c}_j$ for the signal coming from $i$'th RAU. Considering the fact that the distance between any pair of RAU locations is much higher than $d_{min}$ for the GDAS, and using the fact that $\phi(d) = d^{-\alpha}$ is exponentially decreasing function ($2 \le \alpha \le 6$) yields the result that the matrix $\mathbf{A}$ in (28) is a strongly diagonally dominant matrix. This means that matrix $\mathbf{A}$ in (28) is always invertible. If the unique solution of the linear system (27) satisfies also the transmit power constraints, i.e., $\|\mathbf{p}\| > \mathbf{0}$ and $\|\mathbf{p}\|_1 \le p_{tx}^{sum}$, then the optimum transmit power vector, denoted by $\mathbf{p}_{cls}^{opt}$, gives

$$\mu_k^\theta = \bar{\theta}_a(\mathbf{p}_{cls}^{opt}; \mathbf{c}_k) + \delta/\upsilon \quad (29)$$

where $k = 1, \cdots, K$. On the other hand, if the unique solution of (27) does not satisfy the sum power constraint, then the $\mathbf{p}_{cls}^{opt}$ may be found by the following constrained linear optimization

$$\mathbf{p}_{cls}^{opt} = \min_{\mathbf{p}} \left\{ \sum_{k=1}^{K} \left| \bar{\theta}_a(\mathbf{p}; \mathbf{c}_k) + \delta/\upsilon - \mu_k^\theta \right| \right\} \quad (30)$$

s.t. $\|\mathbf{p}\| > \mathbf{0}$ and $\|\mathbf{p}\|_1 \le p_{tx}^{sum}$

Let's denote the globally optimum power vector for the RAU location matrix $\mathbf{C}_{cls} = [\boldsymbol{\mu}_1^x \; \boldsymbol{\mu}_2^x \; \cdots \; \boldsymbol{\mu}_K^x]^T$ as $\mathbf{p}_{glob}^{opt} = [p_1^{opt} \; p_2^{opt} \; \cdots \; p_K^{opt}]^T$, which can be found by any constrained linear optimization techniques minimizing the ergodic capacity error in (14) over the complete location area $\mathbf{x} \in \overline{\Omega}$:

$$\mathbf{p}_{glob}^{opt} = \min_{\mathbf{p}} \{ J_{E,1}(\mathbf{p}, \mathbf{C}) \} \quad (31)$$

s.t. $\|\mathbf{p}\| > \mathbf{0}$ and $\|\mathbf{p}\|_1 \le p_{tx}^{sum}$

From (24), (30), and (31), we conclude that

$$J_{E,1}(\mathbf{p}_{glob}^{opt}, \mathbf{C}_{cls}) \le J_{E,1}(\mathbf{p}_{cls}^{opt}, \mathbf{C}_{cls}) \le UB_{E,1}(\mathbf{p}_{cls}^{opt}, \mathbf{C}_{cls}) \quad (32)$$

where $\mathbf{p}_{cls}^{opt}$ and $\mathbf{p}_{glob}^{opt}$ are obtained by (30), and (31). Comparing (25) with (26) and choosing $\omega_1 = \upsilon v_{glob}$ and $\omega_2 = \upsilon$ in $\mathbf{x}_a = [\omega_1 \mathbf{x}^T\ \omega_2 \bar{\theta}_d(\mathbf{x})]^T$ in (26), we conclude from (25), (26), and (29)-(30) that

$$UB_{E,1}(\mathbf{p}_{cls}^{opt}, \mathbf{C}_{cls}) = Q_{E,1}(\boldsymbol{\mu}_1^{I/O}, \cdots, \boldsymbol{\mu}_K^{I/O}) \qquad (33)$$

Eq.(33) implies that in order to minimize the UB of the ergodic capacity error for the case of *globally optimum* transmit power over the whole GDAS area $\bar{\Omega}$, we can optimally determine the RAU locations simply by the codebook design process in (26) in *location-SNR/capacity* space. Because we assumed that the codebook design process in (26) is optimum (the codebook design optimization is out of the scope of this paper), then increasing the number of RAUs decreases the upper bound $UB_1(\mathbf{p}_{cls}^{opt}, \mathbf{C}_{cls})$, which completes the proof.

2) *Location Optimization using $J_{E,2}(\mathbf{C})$ in (15)*

In this subsection, we derive an UB to the cell averaged ergodic capacity error squared in (15): Following the steps from (17) to (25) for $J_{E,2}(\mathbf{C})$ in (15) instead of $J_{E,1}(\mathbf{C})$ in (14) and using the fact that the $l_1$-norm of a $(q+1)$-dimensional vector is not greater than $\sqrt{q+1}$ times its $l_2$-norm yields the following UB

$$J_2(\mathbf{p},\mathbf{C}) = E_{\mathbf{x}} E_{h,s} \{(\log_2(1+\theta_d(\mathbf{x})+\delta) - \log_2(1+\theta_a(\mathbf{C},\mathbf{x})))^2\} \qquad (34)$$
$$\leq UB_2(\mathbf{p},\mathbf{C})$$

where

$$UB_{E,2}(\mathbf{p},\mathbf{C}) = (q+1)\upsilon^2 v_{glob}^2 \int_{\mathbf{x}\in\bar{\Omega}} \|\mathbf{x} - \mathbf{c}_{k(\mathbf{x})}\|_2^2 f(\mathbf{x})d\mathbf{x} \qquad (35)$$
$$+ (q+1)\upsilon^2 \int_{\mathbf{x}\in\bar{\Omega}} (\bar{\theta}_d(\mathbf{x}) + \delta/\upsilon - \bar{\theta}_a(\mathbf{c}_{k(\mathbf{x})}))^2 f(\mathbf{x})d\mathbf{x}$$

where $\mathbf{x} \in \Re^k$, and $\upsilon$ is the Lipschitz constant of the function $\log_2(\cdot)$ in area $\bar{\Omega}$, and $v_{glob} = \alpha\gamma\left(\sum_{k=1}^{K} p_k \bar{s}_k\right)/\left(\sigma_\varsigma^2 d_{\min}^{(\alpha+1)}\right)$. We call the upper bound $UB_{E,2}(\lambda,\mathbf{C})$ in (25) (which is to be minimized) as "IOCC criterion" for optimizing the locations of the RAUs of DAS. Let us consider a codebook design process in vector quantization using a clustering algorithm which minimizes the following quantization squared

error

$$Q_{E,2}(\mu_1^{I/O},\cdots,\mu_K^{I/O}) = E\left\{\left\|x^{I/O} - \mu_{k(x^{I/O})}^{I/O}\right\|_2^2\right\} \quad (36)$$

$$= \int_{x\in\Omega} \left(\omega_1\left\|x - \mu_{k(x^{I/O})}^x\right\|_2^2 + \omega_2\left(\bar{\theta}_d(x) + \delta/\upsilon - \mu_{k(x_a)}^\theta\right)^2\right) f(x)dx$$

where $\bar{\theta}_d(\mathbf{x})$ is the location-dependent desired SNR function, and $\mathbf{x}^{I/O} = [\omega_1\mathbf{x}^T \ \omega_2(\bar{\theta}_d(\mathbf{x})+\delta)]^T$ as defined before in (26).

*Proposition 2*: Considering the optimum transmit power vector $\mathbf{p}_{glob}^{opt}$, i.e., $J_{E,2}(\mathbf{p}_{glob}^{opt},\mathbf{C}) \leq J_{E,2}(\mathbf{p},\mathbf{C})$ for any **p**, the UB $UB_{E,2}(\mathbf{p}=\mathbf{p}_{glob}^{opt},\mathbf{C})$ in (35) to the ergodic capacity error squared $J_{E,2}(\mathbf{p}_{glob}^{opt},\mathbf{C})$ can be decreased by a clustering process (like Lloyd, k-means algorithms) in *location-SNR/capacity* space.

*Proof:* Following the steps from (26) to (32) for the $J_{E,2}(\mathbf{p},\mathbf{C})$ in (15) and $Q_{E,2}(\{\mu_k^{I/O}\}_{k=1}^K)$ in (36), we conclude that

$$J_{E,2}(\mathbf{p}_{glob}^{opt},\mathbf{C}_{cls}) \leq J_{E,2}(\mathbf{p}_{cls}^{opt},\mathbf{C}_{cls}) \leq UB_{E,2}(\mathbf{p}_{cls}^{opt},\mathbf{C}_{cls}) \quad (37)$$

where $\mathbf{p}_{cls}^{opt}$ is optimum only for the set of $\{\mathbf{c}_k = \mu_k^x, \bar{\theta}_a = \mu_k^\theta\}_{k=1}^K$, and $\mathbf{p}_{glob}^{opt} = \min_\mathbf{p}\{J_{E,2}(\mathbf{p},\mathbf{C})\}$ such that $\|\mathbf{p}\| > \mathbf{0}$ and $\|\mathbf{p}\|_1 \leq p_{tx}^{sum}$.

Because matrix **A** is non-singular, if the unique solution of the linear system (27) satisfies also the transmit power constraints, then the optimum transmit power vector, denoted by $\mathbf{p}_{cls}^{opt}$ satisfies $\mu_k^\theta = \bar{\theta}_a(\mathbf{p}_{cls}^{opt};\mathbf{c}_k) + \delta/\upsilon$, for $k=1,...,K$. Othewise the $\mathbf{p}_{cls}^{opt}$ may be found by the following constrained linear squares optimization

$$\mathbf{p}_{cls}^{opt} = \min_\mathbf{p}\left\{\sum_{k=1}^K \left(\bar{\theta}_a(\mathbf{p};\mathbf{c}_k) + \delta/\upsilon - \mu_k^\theta\right)^2\right\} \quad (38)$$

s.t. $\|\mathbf{p}\| > \mathbf{0}$ and $\|\mathbf{p}\|_1 \leq p_{tx}^{sum}$

Using (26), (35) and (37) and choosing $\omega_1 = (q+1)\upsilon^2 v_{glob}^2$ and $\omega_2 = (q+1)\upsilon^2$ in $\mathbf{x}^{I/O} = [\omega_1\mathbf{x}^T \ \omega_2(\bar{\theta}_d(\mathbf{x})+\delta/\upsilon)]^T$, we

observe that

$$UB_{E,2}\left(\mathbf{p}_{cls}^{opt}, \mathbf{C}_{cls}\right) = Q_{E,2}(\mathbf{\mu}_1^{I/O}, \cdots, \mathbf{\mu}_K^{I/O}) \quad (39)$$

Traditional clustering algorithms like Lloyd, k-means minimizes error squares as in (36) instead of the absolute value of errors as in (26). From (35), (37) and (39), the upper bound $UB_{E,2}\left(\mathbf{p} = \mathbf{p}_{glob}^{opt}, \mathbf{C}\right)$ in (35) can be decreased by a clustering process like Lloyd, k-means in *location-SNR/capacity* space, which completes the proof.

On the other hand, the SDC criterion in [15] (and in all other related papers mentioned in section I) is equal to

$$\begin{aligned}SDC &= E\left\{\left\|\mathbf{x} - \mathbf{c}_{k(\mathbf{x})}\right\|_2^2\right\} \\ &= \int_{\mathbf{x} \in \Omega} \left\|\mathbf{x} - \mathbf{c}_{k(\mathbf{x})}\right\|_2^2 f(\mathbf{x}) d\mathbf{x}\end{aligned} \quad (40)$$

which is mathematically equal to eq.(12) in [15]. Comparing (25) and (40), we see that minimizing only the first integral of the IOCC is equal to minimizing the SDC in [15]. Thus, IOCC includes SDC as a special case. The first integral is related to the input space (i.e., location space), while the second integral is related to the output space (i.e., SNR/capacity space). So, the SDC considers only location space. In this paper, we device an IOCC based RAU location algorithm which takes the both integrals in input-output space into account when minimizing the UB.

As far as the UBs are concerned, using the steps (19)-(32), and the norm properties of vectors, we can obtain a tighter UB than (35) as follows: Defining

$$\begin{aligned}d_{1,k} &= \left\|\mathbf{x} - \mathbf{c}_k\right\|_2 \\ d_{2,k} &= \left\|\mathbf{c}_{k(\mathbf{x})} - \mathbf{c}_k\right\|_2 \\ d_{\min,k} &= \min(d_{1,k}, d_{2,k}), \text{ and} \\ v_{k,\mathbf{x}} &= \frac{1}{\sigma_\varsigma^2} \sum_{n=1}^K \lambda_n \frac{\alpha}{d_{\min,n}^{\alpha+1}}\end{aligned} \quad (41)$$

we obtain

$$UB_{E2,opt}(\mathbf{p}_{cls}^{opt}, \mathbf{C}) = \quad (42)$$
$$\upsilon^2 \int_{\mathbf{x}\in\bar{\Omega}} \left( \nu_{k,\mathbf{x}} \|\mathbf{x} - \mathbf{c}_{k(\mathbf{x})}\|_2 + |\bar{\theta}_d(\mathbf{x}) - \bar{\theta}_a(\mathbf{C}, \mathbf{c}_{k(\mathbf{x})})| \right)^2 f(\mathbf{x}) d\mathbf{x}$$

### B. IOCC in Deterministic Setting

In Section III.A, we analyze the RAU location problem from a statistical point of view. In what follows, we derive similar results in a deterministic setting in order to devise the RAU allocation algorithm: Let's assume that we are given $L$ location samples from the user distribution $f(\mathbf{x})$, denoted by set $\{\mathbf{x}_l\}_{l=1}^L$, and corresponding location-dependent desired ergodic capacity function, denoted by $\{\Gamma_d(\mathbf{x}_l)\}_{l=1}^L$. Then the squared ergodic capacity error squared in (15) is approximated by these $L$ samples as follows:

$$J_{E,2}(\mathbf{p}, \mathbf{C}) \approx J_2(\mathbf{p}, \mathbf{C}) = \frac{1}{L} \sum_{l=1}^{L} (\Gamma_d(\mathbf{x}_l) - \Gamma_a(\mathbf{x}_l))^2 \quad (43)$$

where $\Gamma_d(\mathbf{x}_l)$ and $\Gamma_a(\mathbf{x}_l)$ is the desired/demanded and actual/supplied ergodic capacity, respectively, for given location $\mathbf{x}_l$. Following the steps (16) to (32) in a deterministic setting, we similarly obtain the upper bound $UB_2(\mathbf{p}_{glob}^{opt}, \mathbf{C})$ to the ergodic capacity error in (43) with the globally optimum transmit power as follows:

$$J_2(\mathbf{p}_{glob}^{opt}, \mathbf{C}) \leq UB_2(\mathbf{p}_{glob}^{opt}, \mathbf{C}), \quad (44)$$
$$UB_2(\mathbf{p}_{glob}^{opt}, \mathbf{C}) = \frac{(q+1)}{L} \sum_{l=1}^{L} \|\upsilon \nu_{cls}^{opt}(\mathbf{x}_l - \mathbf{\mu}_{k(l)}^x)\|_2^2$$
$$+ \frac{(q+1)\upsilon^2}{L} (\bar{\theta}_d(\mathbf{x}_l) + \delta/\upsilon - \bar{\theta}_a(\mathbf{C}, \mathbf{c}_{k(l)}))^2$$

where $\mathbf{p}_{glob}^{opt}$ is *global* optimum transmit power for the complete set $\{\mathbf{x}_l, \theta_d(\mathbf{x}_l)\}_{l=1}^L$, and $\mathbf{p}_{cls}^{opt}$ is optimum only for the set of codebook vectors $\{\mathbf{\mu}_k^x, \mu_k^\theta\}_{k=1}^K$ and $k(l)$ represents the index of the corresponding codebook vector in *location-SNR/capacity* space. The quantization squared error in (36) is approximated by

$$Q_{E,2}(\{\mathbf{\mu}_k^{I/O}\}_{k=1}^K) \approx Q_2(\{\mathbf{\mu}_k^{I/O}\}_{k=1}^K) = \frac{1}{L} \sum_{l=1}^{L} \|\mathbf{x}_l^{I/O} - \mathbf{\mu}_{k(\mathbf{x}_l^{I/O})}^{I/O}\|_2^2 \quad (45)$$
$$= \omega_1 \|\mathbf{x}_l - \mathbf{\mu}_{k(\mathbf{x}_l^{I/O})}^x\|_2^2 + \omega_2 (\bar{\theta}_d(\mathbf{x}_l) + \delta/\upsilon - \mu_{k(\mathbf{x}_l^{I/O})}^\theta)^2$$

where $\omega_1, \omega_2 > 0$.

*Definition: Feasible DAS:* Because matrix **A** is a diagonally dominant (positive) matrix, it is always non-singular. If the unique solution of the linear system (27) satisfies also the transmit power constraints, i.e., $\|\mathbf{p}\|>0$ and $\|\mathbf{p}\|_1 \leq p_{tx}^{sum}$, then, we define such a system as *feasible DAS*.

So, for feasible DAS, $\mu_k^\theta = \bar{\theta}_a(\mathbf{p}_{cls}^{opt}; \mathbf{c}_k) + \delta/\upsilon$, for $k=1,...,K$.

*Proposition 3*: Let's consider $L$ samples of $\{\mathbf{x}_l\}_{l=1}^L$, and corresponding location-dependent desired average SNRs, denoted by $\{\bar{\theta}_d(\mathbf{x}_l)\}_{l=1}^L$, which is related to the location-dependent desired ergodic capacity $\{\Gamma_d(\mathbf{x}_l)\}_{l=1}^L$. The upper bound $UB_2(\mathbf{p}=\mathbf{p}_{glob}^{opt}, \mathbf{C})$ in (43) to the ergodic capacity error squared $J_2(\mathbf{p}_{glob}^{opt}, \mathbf{C})$ in (44) can be decreased by a clustering process (like Lloyd, *k*-means, etc) in *location-SNR/capacity* space. Provided that the clustering process gives optimum performance, the $J_2(\mathbf{p}_{glob}^{opt}, \mathbf{C})$ can be made arbitrarily small by increasing the number of RAU locations for a *feasible DAS*.

*Proof:* Let's consider the clustering process in (45), which is a quantization function exactly minimized by traditional clustering algorithms like the Lloyd, *k*-means algorithms, etc.

From (44) and (45), choosing $\omega_1 = (q+1)\upsilon^2 v_{glob}^2$ and $\omega_2 = (q+1)\upsilon^2$ in $\mathbf{x}^{I/O} = [\omega_1 \mathbf{x}^T \ \omega_2(\bar{\theta}_d(\mathbf{x})+\delta/\upsilon)]^T$ for a *feasible* DAS gives

$$UB_2(\mathbf{p}_{cls}^{opt}, \mathbf{C}_{cls}) = Q_2(\{\boldsymbol{\mu}_k^{I/O}\}_{k=1}^K) \qquad (46)$$

Therefore, increasing the number of RAUs decreases the upper bound $UB_{E,2}(\mathbf{p}_{cls}^{opt}, \mathbf{C}_{cls})$. Here we assume that the clustering process gives optimum performance because the clustering process itself is out of the scope of this paper. Then, an optimal clustering process may arbitrarily decrease the $Q_2(\{\boldsymbol{\mu}_k^{I/O}\}_{k=1}^K)$ by arbitrarily increasing the number of RAUs, $K$. In the limit case where $K=L$ (i.e., number of samples is equal to the number of clusters), obviously $Q_2(\{\boldsymbol{\mu}_k^{I/O}\}_{k=1}^K)$, which completes the proof.

The IOCC criterion minimizes the upper bound $UB_2(\lambda_{glob}^{opt}, \mathbf{C})$ in (44) in both input (*location*) and output (*SNR/capacity*) space at the same time. Considering the SDC criterion in [15] in the context of the upper bound $UB_2(\lambda_{glob}^{opt}, \mathbf{C})$ in (44), we see that the SDC is equal to minimizing only the first term of the $UB_2(\lambda_{glob}^{opt}, \mathbf{C})$. Therefore, while the RAU locations are determined by codebook design merely in input (*location*) space by the SDC, the RAU locations of the proposed IOCC is determined by not only input (*location*) space but also output (*SNR/capacity*) space, simultaneously. The proposed IOCC based RAU location algorithm for a given set of $L$ samples $\{\mathbf{x}_l, \theta_d(\mathbf{x}_l)\}_{l=1}^{L}$ is presented in Table 1. As explained before, there exists a small number $\varepsilon > 0$ such that $\log_2(1+\bar{\theta}_d(\mathbf{x})) \geq \Gamma_d(\mathbf{x}) \geq \log_2(1+\bar{\theta}_d(\mathbf{x})) - \varepsilon$, and therefore, for a given arbitrary $\Gamma_d(\mathbf{x})$, there exists a corresponding $\bar{\theta}_d(\mathbf{x})$. The weights step 1 is $\omega_1 = (q+1)\upsilon^2 v_{glob}^2$ and $\omega_2 = (q+1)\upsilon^2$ where the variables are defined as above.

**Table 1** IOCC based RAU location algorithm

1. Define weighted and augmented vectors $\mathbf{x}_l^{I/O} = [\omega_1 \mathbf{x}_l^T \; \omega_2 \theta_d(\mathbf{x}_l)]^T$ in input-output space, where $l = 1, \cdots, L$.
2. Using the set $\{\mathbf{x}_l^{I/O}\}_{l=1}^{L}$, find $K$ codebook vectors, denoted by $\{\boldsymbol{\mu}_k^{I/O} = [\omega_1 \boldsymbol{\mu}_k^{xT} \; \omega_2 \mu_k^{\theta}]^T\}_{k=1}^{K}$, using any clustering algorithm like e.g. k-means, Lloyd, etc. in input-output space.
3. Project the augmented codebook vectors $\{\boldsymbol{\mu}_k^{I/O}\}_{k=1}^{K}$ onto the input space (location space), and obtain the RAU locations as $\mathbf{c}_k = \boldsymbol{\mu}_k^x$, $k = 1, \cdots, K$.

As seen from Table 1, all we need to apply the IOCC is the user location distribution function and the location specific desired/demanded ergodic capacity function. In the IOCC, if $\omega_2 = 0$ or $\omega_2$ is a constant, then the SNR/capacity space has no impact on the RAU locations, and in this case IOCC reduces to the SDC in [15] exactly.

We'd like to note that as in the UB in (42), we can obtain tighter UB (denoted as $UB_{opt}(\mathbf{p}_{glob}^{opt}, \mathbf{C})$) for the deterministic case as follows:

$$UB_{opt}(\lambda_{glob}^{opt}, \mathbf{C}) = \left( \frac{\upsilon^2}{L} \sum_{k=1}^{K} v_{k,\mathbf{x}_l} \|\mathbf{x}_l - \mathbf{c}_{k(l)}\|_2 + |\theta_d(\mathbf{x}_l) + \delta/\upsilon - \theta_a(\mathbf{C}, \mathbf{c}_{k(l)})| \right)^2 \quad (47)$$

where $v_{k,\mathbf{x}_l}$ is defined as in (41).

## IV. SIMULATION RESULTS

Without loss of generality, a direct-sequence (W)CDMA/TDD wireless network is considered in all examples of the GDAS. For link gain modeling, pathloss attenuation factor $2 \leq \alpha \leq 6$, the log-normally distributed $s_{ij}$ in eq. is generated according to the model in [26], and the lognormal variance is 6 dB.

### A. Simulation Results with transmit power control

In this part, there is optimal transmit power control, and the sum of the RAU transmit powers is not greater than 1 W for all simulations, i.e. $\|\mathbf{p}\|_1 \leq 1 [W]$.

*Example 1:* In this example, we examine linear cell case scenario [18]. One-dimensional cell has 2km length. The location-specific desired ergodic capacity linearly reduces from 5.35 [bps/Hz] at *x=0* m to 3.45 [bps/Hz] at *x=2000* m. The user locations are evenly distributed over the lineer DAS, and we take 100 samples accordingly $\{x_l, \Gamma_d(x_l)\}_{l=1}^{100}$ to demonstrate how the upper bounds and capacity errors evolve in a deterministic setting as presented in section III.B. The normalized upper bounds $UB_1$, $UB_2$ and $UB_{opt}$ and the mean squared error $J(\lambda_{glob}^{opt}, \mathbf{C})$ in (43) by the IOCC, denoted by $J_{IOCC}$, in logarithmic scale with respect to the number of RAUs are given in Fig. 1. All the logarithmic scale values are normalized by the 10-log of the maximum of $UB_2$. The Fig. 1 confirms the findings in section III.B: The derived UBs decrease as the number of RAUs increases (because the more RAUs, the less the quantization error in input-output space), and the UBs to the squared SNR error can be made *arbitrarily* small by increasing the number of RAUs.

Normalized mean squared ergodic capacity error with respect to the number of RAUs for the SDC and IOCC, denoted by $J_{SDC}$ and $J_{IOCC}$, respectively, are shown in Fig. 2. The values in Fig. 2 are normalized by the maximum of $J_{SDC}$ in linear scale. The figure shows that the IOCC outperforms the SDC in terms of the mean squared error in capacity.

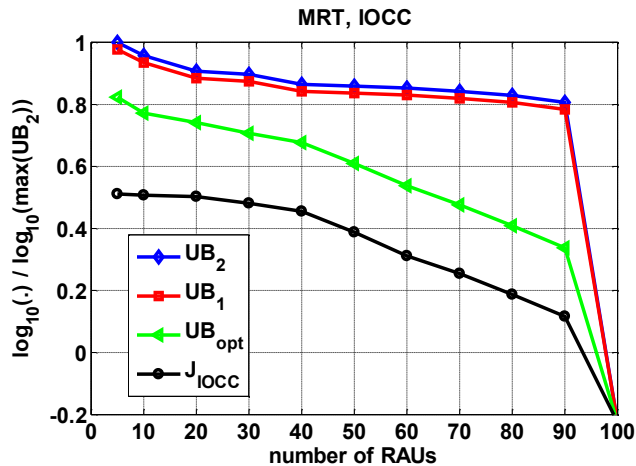

**Fig. 1.** UBs and mean squared error in Example 1 for $L=100$.

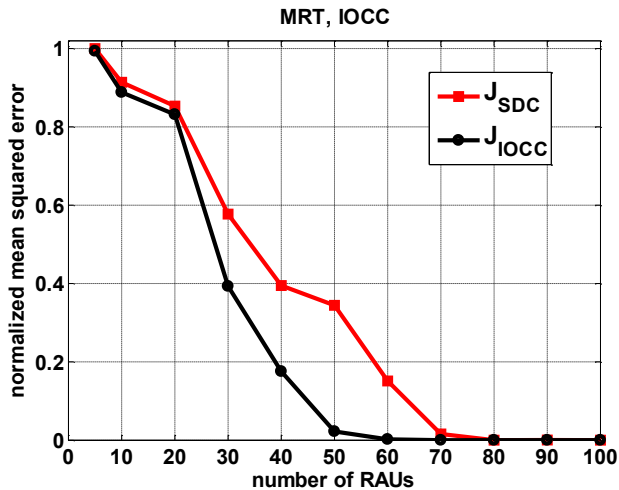

**Fig. 2** Normalized mean squared SNR error with respect to the number of RAUs for the SDC and IOCC ($L=100$).

*Example 2*: In this example, we examine a two-dimensional general DAS scenario as in [15]. The user locations are drawn from a PPP process whose density is 0.003. The location-specific desired ergodic capacity function linearly reduces with respect to the distance from 5.35 [bps/Hz] at $\mathbf{x}=[0\ 0]^T$ to 3.45 [bps/Hz] at $\mathbf{x}=[500\ 500]^T$. In order to give an insight into the difference between the RAU locations found by the SDC and the proposed IOCC, we present the RAU locations for the case $N=2$ and $N=5$ in Fig. 3 and Fig. 5, respectively, for the same snapshot. Standard k-means algorithm is used for the both SDC and IOCC, and different colors represent different clusters found by the clustering algorithm. The RAU locations of both the SDC and the IOCC are indicated on the same plot in Fig. 4 for the same snapshot when $N=5$. Examining Fig. 3, Fig. 4 and Fig. 5, it's observed that the RAU locations found by the IOCC (circles in red) get closer to the areas where the desired SNRs are higher, as compared to those by the SDC (diamonds in blue). This is because the locations of the RAUs are determined not only by the MS location distribution but by the location-specific target-SNRs also. So, Examining Fig. 3, Fig. 4 and Fig. 5 confirm the findings in section III.

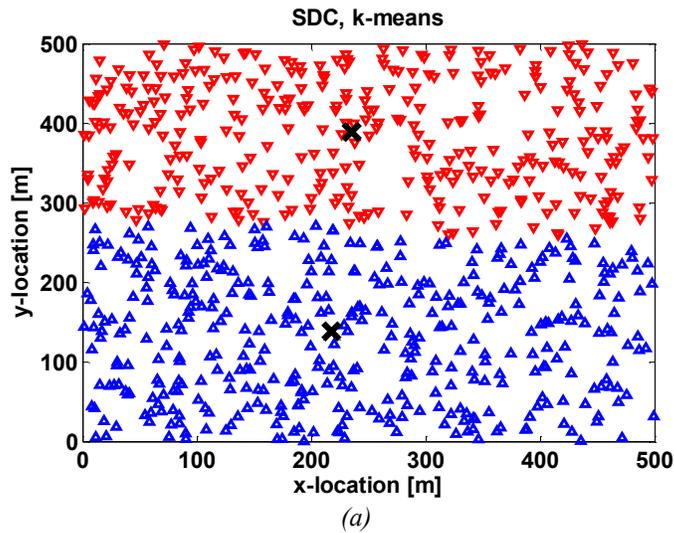

(a)

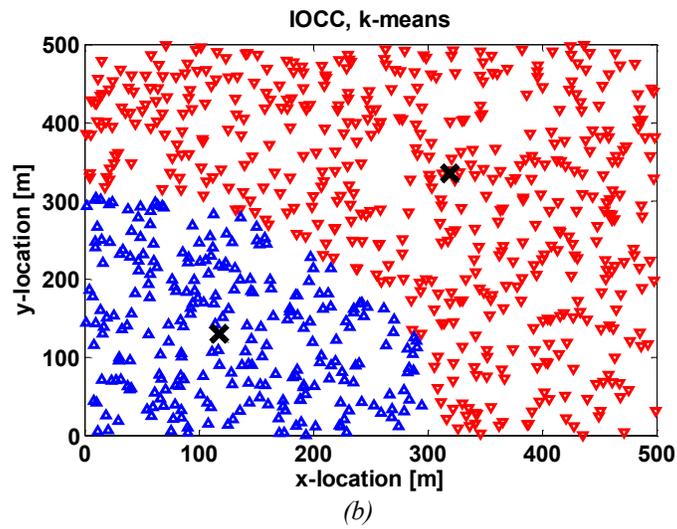

*(b)*

**Fig. 3** The RAU locations denoted by **x** and clusters found by *(a)* SDC, and *(b)* IOCC for *N=2*.

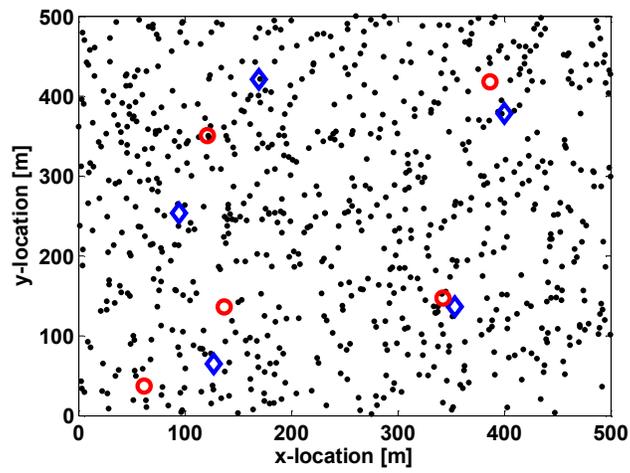

**Fig. 4** The RAU locations by the SDC (blue diamond) and IOCC (red circle) for *N=5*.

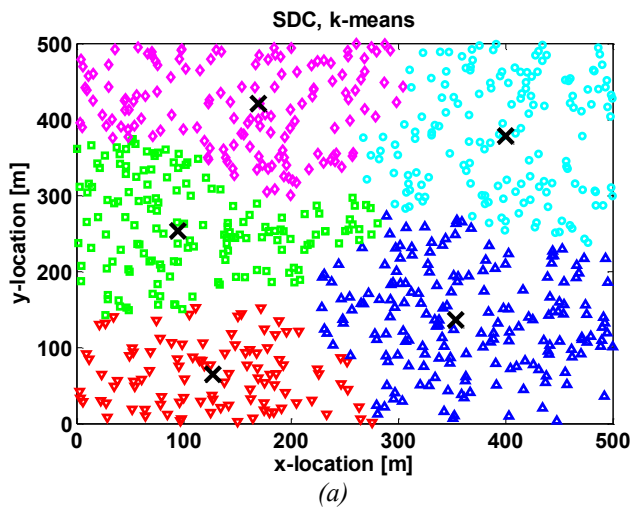

*(a)*

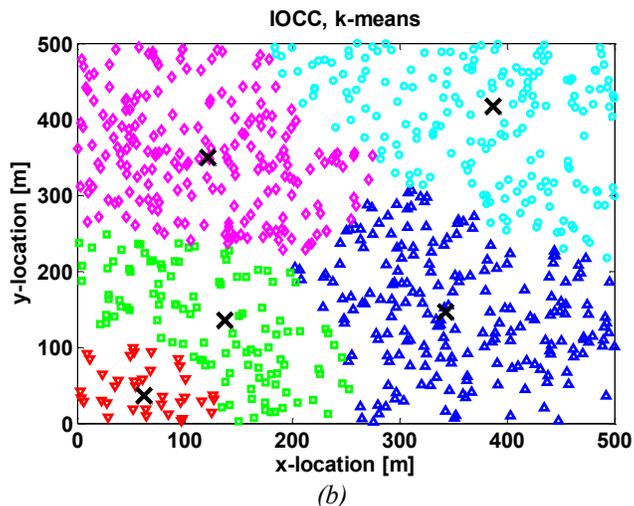

**Fig. 5** The RAU locations denoted by **x** and clusters found by *(a)* SDC, and *(b)* IOCC for *N=5*.

The normalized upper bounds $UB_1$, $UB_2$ and $UB_{opt}$ and the mean squared error $J_2(\lambda_{glob}^{opt}, \mathbf{C})$ in (15), (43) by the IOCC, denoted by $J_{IOCC}$, in logarithmic scale with respect to the number of RAUs are given in Fig. 6. The figure confirms the findings in section III: The derived UBs decrease as the number of RAUs increases (because the more RAUs, the less the quantization error in input-output space). Normalized mean squared SNR error with respect to the number of RAUs for the SDC and IOCC, denoted by $J_{SDC}$ and $J_{IOCC}$, respectively, are shown in Fig. 7. The values in Fig. 7 are normalized by the maximum of $J_{SDC}$ in linear scale. The figure shows that the IOCC outperforms the SDC in terms of the mean squared SNR error in ergodic capacity.

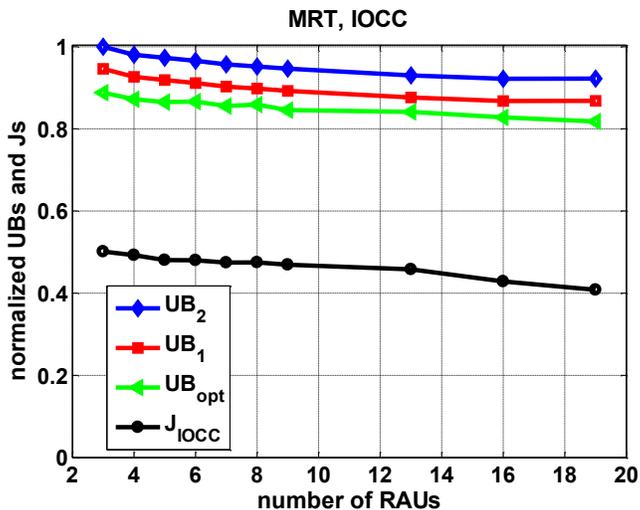

**Fig. 6** UBs and mean squared error in Example 2.

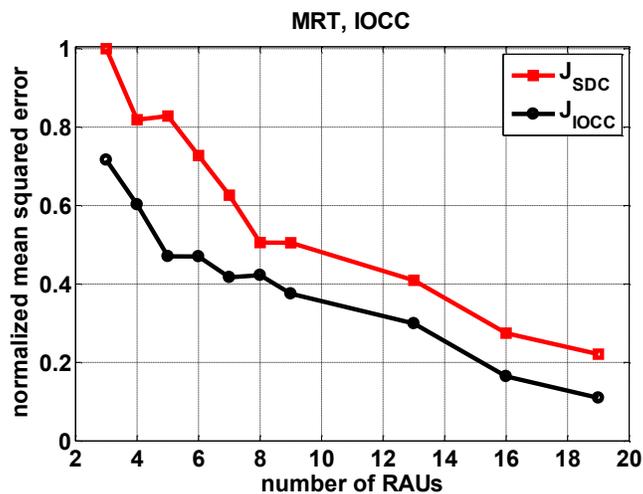

**Fig. 7** Normalized mean squared SNR error with respect to the number of RAUS for the SDC and IOCC.

*B. Simulation Results without transmit power control*

In this section, there is no transmit power control, and the total DAS transmit power is fixed, and is equally distributed over the RAUs. The main goal of this part is to compare the IOCC and SDC performances in terms of the cell averaged ergodic capacity for a DAS with equal and fixed RAU transmit power. The defined "wasted capacity" concept is shown in the schema in Fig. 8 where target capacity (with $\delta=0$), and SDC (red) and IOCC (blue) capacities are sketched using a simple pathloss-based model just for illustration purposes for *K=2* in a linear DAS. The RAU locations with the SDC are the IOCC are shown by red and blue circles, respectively. In what follows, we compare the performances of the SDC and the IOCC in terms of the cell averaged ergodic capacity $E_\mathbf{x}\{\Gamma_a(\mathbf{x})\}$ as well as the cell averaged *effective* ergodic capacity $E_\mathbf{x}\{\Gamma_{\mathit{eff}}(\mathbf{x})\}$.

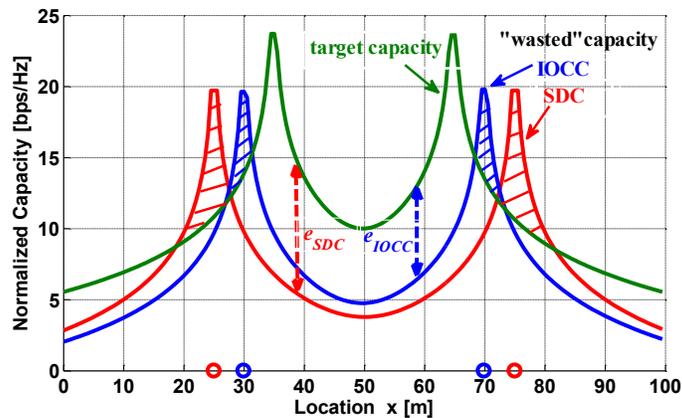

**Fig. 8** The defined "wasted capacity" concept.

*Example 3*: In this example, we examine a two-dimensional general DAS scenario as in [15]. The MS locations are drawn from a PPP process whose density is 0.02. The location-specific desired ergodic capacity function linearly reduces with respect to the distance from 9.65 [bps/Hz] at $\mathbf{x}=[250\ 250]^T$ to 3.45 [bps/Hz] at the DAS borders. The sum of the RAU transmit powers is equal to 1 W, i.e., $\|\mathbf{p}\|_1 = 1\,[W]$, which is evenly distributed over the RAUs for all simulations such that $p_1 = \cdots = p_K = 1/K\ [W]$.

The cell averaged *effective* ergodic capacity $E_{\mathbf{x}}\{\Gamma_{\mathit{eff}}(\mathbf{x})\}$ for the SDC and IOCC for pathloss exponent $\alpha = 3$, and $\alpha = 4$ are shown in Fig. 9 and Fig. 10, respectively. The cell averaged ergodic capacity $E_{\mathbf{x}}\{\Gamma_a(\mathbf{x})\}$ results are presented in Fig. 11.

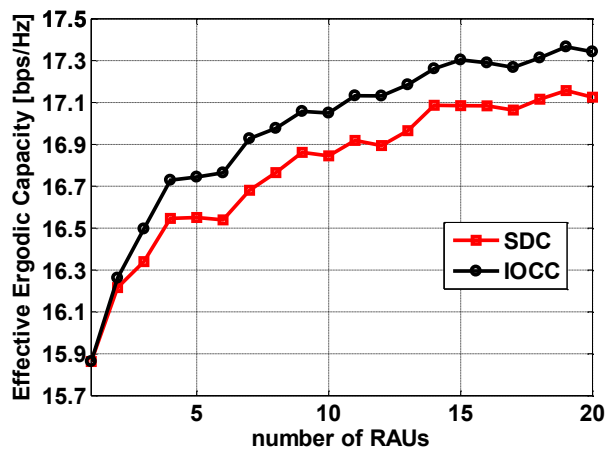

**Fig. 9** The cell averaged *effective* ergodic capacity for $\alpha = 3$ in Example 3.

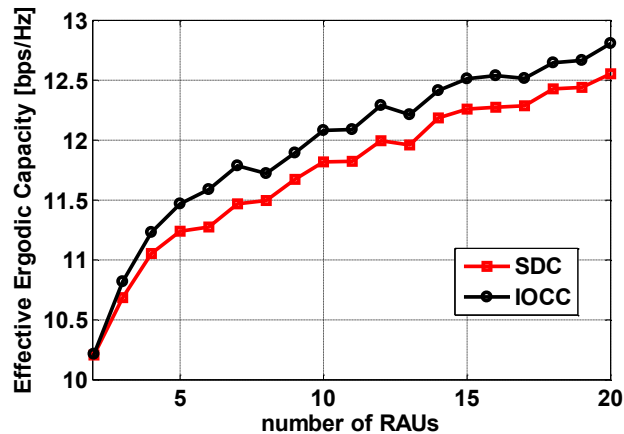

**Fig. 10** The cell averaged *effective* ergodic capacity for $\alpha=4$ in Example 3.

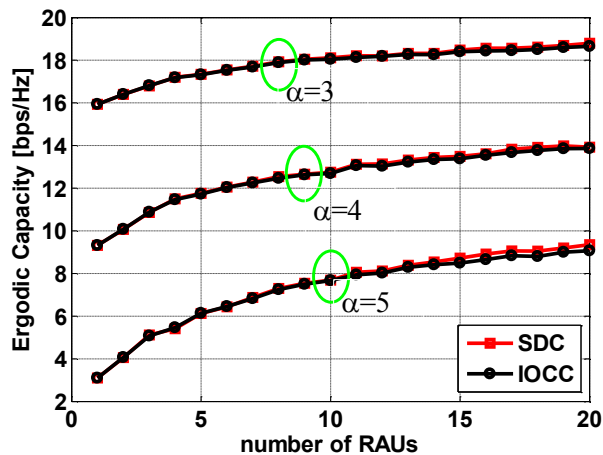

**Fig. 11** The cell averaged ergodic capacity for different pathloss exponent values in Example 3.

The results in Fig. 9 and Fig. 10 show that the IOCC outperforms the SDC in terms of the *effective* ergodic capacity, where the gain by the IOCC is around 0.2 [bps/Hz]. The cell averaged ergodic capacity results are presented in Fig. 11. The results in Fig. 11 suggest that the SDC performance is either comparable to or very slightly better than those of the IOCC. So, from Fig. 9, Fig. 10, and Fig. 11, a clear *effective* ergodic capacity gain is obtained by the IOCC at the cost of the possibility of a slight decrease in the cell-averaged ergodic capacity.

## V. Conclusions

In this paper, we analyze the following question: For a given *arbitrary* user distribution, and location-dependent desired ergodic capacity function, which is *arbitrary*, what are the optimum RAU locations of the GDAS minimizing the cell averaged ergodic capacity error, and thus maximizing the *effective* ergodic capacity? We propose a novel criterion, called IOCC, for the RAU location optimization. Our investigations show that

*i)* the IOCC provides an upper bound to the cell averaged ergodic capacity error;

*ii)* the derived upper bound is equal to a weighted quantization error function in *location-capacity* space (input-output space) and

*iii)* the upper bound can be made arbitrarily small by a clustering process increasing the number of RAUs with optimal transmit power control for a *feasible* system.

The IOCC converts the RAU location problem into a codebook design problem in vector quantization in *input-output* space, and thus includes the Squared Distance Criterion (SDC) for DAS in [15] (and other related papers) as a special case, which takes only the *input* space into account. Computer simulations confirm the theoretical findings and show that the IOCC outperforms the SDC for GDAS both in terms of the cell averaged capacity *error*, and in terms of the defined cell averaged *"effective"* ergodic capacity at the cost of a probable slight decrease in the cell-averaged ergodic capacity.

## APPENDIX

In what follows, we examine the global Lipschitz constant of the average SNR function $\bar{\theta}_a(\cdot)$ in (16) for the interval $[d_{min}, \infty)$ where $d_{min}$ is the minimum distance between the user and any RAU: We first show that the path loss function $\phi(d) = d^{-\alpha}$ for the interval $[d_{min}, \infty)$ has the Lipschitz constant as $\vartheta = \alpha / d_{min}^{(\alpha+1)}$, where $\alpha$ is the path loss exponent: Because $\phi(d) = d^{-\alpha}$ is a differentiable function in $[d_{min}, \infty)$, we can apply the mean value theorem $\forall d_i, d_j \in [d_{min}, \infty)$ as follows:

$$\phi(d_i) - \phi(d_j) = (d_i^{-\alpha} - d_j^{-\alpha}) \phi'(\mu d_i + (1-\mu) d_j) \quad (48)$$

where $\mu \in [0, 1]$. The derivative of $\phi(d)$ is $\phi'(d) = -\alpha d^{-(\alpha+1)}$. So, the absolute value of the derivative for the interval $[d_{min}, \infty)$ is at $d = d_{min}$. Thus,

$$|\phi(d_i) - \phi(d_j)| \leq \vartheta |d_i - d_j| \qquad (49)$$

where $\vartheta = \alpha / d_{min}^{(\alpha+1)}$ is the Lipschitz constant of the path loss function $\phi(d)$.

It's assumed that large-scale and small-scale fading random variables $s_n$ and $g_{n,\mathbf{x}}$ are independent, and the average large-scale fading $E_s\{s_n\}$ is RAU location-specific. Denoting the average small-scale fadings at locations $\mathbf{x}_i$ and $\mathbf{x}_j$ as $\bar{g}_{n,\mathbf{x}_i} = E_h\{g_n(\mathbf{x}_i)\}$ and $\bar{g}_{n,\mathbf{x}_j} = E_h\{g_n(\mathbf{x}_j)\}$, respectively, we define

$$\gamma = \left| \frac{\max\left\{\bar{g}_{n,\mathbf{x}}\|\mathbf{x}_i - \mathbf{c}_k\|_2^{-\alpha}, \bar{g}_{n,\mathbf{c}_k}\|\mathbf{x}_j - \mathbf{c}_k\|_2^{-\alpha}\right\}}{\|\mathbf{x}_i - \mathbf{c}_k\|_2^{-\alpha} - \|\mathbf{x}_j - \mathbf{c}_k\|_2^{-\alpha}} \right| \qquad (50)$$

Using (50) and the fact that the pathloss function $\phi(d) = d^{-\alpha}$ is a decreasing function, we observe that

$$\left| \bar{g}_{n,\mathbf{x}_i} \|\mathbf{x}_1 - \mathbf{c}_k\|_2^{-\alpha} - \bar{g}_{n,\mathbf{x}_j} \|\mathbf{x}_2 - \mathbf{c}_k\|_2^{-\alpha} \right| \qquad (51)$$
$$\leq \gamma \left| \|\mathbf{x}_1 - \mathbf{c}_k\|_2^{-\alpha} - \|\mathbf{x}_2 - \mathbf{c}_k\|_2^{-\alpha} \right|$$

From the average SNR function $\bar{\theta}_a(\cdot)$ in (16), we have

$$\|\bar{\theta}_a(\mathbf{x}_i) - \bar{\theta}_a(\mathbf{x}_j)\|_1 = \frac{1}{\sigma_\varsigma^2} \sum_{k=1}^{K} \bar{s}_n \left\| \bar{g}_{k,\mathbf{x}_i} \phi(\|\mathbf{x}_i - \mathbf{c}_k\|_2) - \bar{g}_{k,\mathbf{x}_j} \phi(\|\mathbf{x}_j - \mathbf{c}_k\|_2) \right\|_1 \qquad (52)$$

Using (49),(51), and applying the triangular rule and the definition of the $l_1$-norm of a vector, we obtain

$$\|\bar{\theta}_a(\mathbf{x}_i) - \bar{\theta}_a(\mathbf{x}_j)\|_1 \leq v_{glob} \|\mathbf{x}_i - \mathbf{x}_j\|_2, \quad \forall \mathbf{x}_i, \mathbf{x}_j \in \overline{\Omega} \qquad (53)$$

where $v_{glob} = \alpha \gamma \left( \sum_{k=1}^{K} p_k \bar{s}_k \right) / \left( \sigma_\varsigma^2 d_{min}^{(\alpha+1)} \right)$, in which $\alpha$ is the path loss exponent, $\gamma$ is related to the average small-scale fading as defined in (50), $p_k$ is the transmit power of the $k$'th RAU, $\bar{s}_k$ is the average large-scale fading coefficient related to the $k$'th RAU, $\sigma_\varsigma^2$ is the average noise power, and $d_{min}$ is the minimum distance between user location and any RAU. Eq.(53) shows that the average SNR function $\bar{\theta}_a(\cdot)$ in (16) has a global Lipschitz constant $v_{glob}$ for the interval $[d_{min}, \infty)$.